\documentclass{article}

\usepackage{microtype}
\usepackage{graphicx}
\usepackage{subfigure}
\usepackage{booktabs} 

\usepackage{hyperref}

\usepackage[accepted]{icml2024}

\usepackage{amsmath}
\usepackage{amssymb}
\usepackage{mathtools}
\usepackage{amsthm}

\usepackage[capitalize,noabbrev]{cleveref}

\theoremstyle{plain}

\theoremstyle{definition}

\theoremstyle{remark}

\usepackage[textsize=tiny]{todonotes}

\icmltitlerunning{Adversarially Robust Feature Learning for Breast Cancer Diagnosis}

\begin{document}

\twocolumn[
\icmltitle{Adversarially Robust Feature Learning for Breast Cancer Diagnosis}



\icmlsetsymbol{equal}{*}

\begin{icmlauthorlist}
\icmlauthor{Degan Hao}{isp}
\icmlauthor{Dooman Arefan}{rad}
\icmlauthor{Margarita Zuley}{rad}
\icmlauthor{Wendie Berg}{rad}
\icmlauthor{Shandong Wu}{isp,rad,bio,dbmi}
\end{icmlauthorlist}

\icmlaffiliation{isp}{Intelligent Systems Program, University of Pittsburgh, Pittsburgh, USA}
\icmlaffiliation{rad}{Department of Radiology, University of Pittsburgh, Pittsburgh, USA}
\icmlaffiliation{bio}{Department of Bioengineering, University of Pittsburgh, Pittsburgh, USA}
\icmlaffiliation{dbmi}{Department of Biomedical Informatics, University of Pittsburgh, Pittsburgh, USA}
\icmlcorrespondingauthor{Shandong Wu}{wus3@upmc.edu}

\icmlkeywords{Machine Learning, ICML}

\vskip 0.3in
]



\printAffiliationsAndNotice{}  

\begin{abstract}
Adversarial data can lead to malfunction of deep learning applications. It is essential to develop deep learning models that are robust to adversarial data while accurate on standard, clean data. In this study, we proposed a novel adversarially robust feature learning (ARFL) method for a real-world application of breast cancer diagnosis. ARFL facilitates adversarial training using both standard data and adversarial data, where a feature correlation measure is incorporated as an objective function to encourage learning of robust features and restrain spurious features. To show the effects of ARFL in breast cancer diagnosis, we built and evaluated diagnosis models using two independent clinically collected breast imaging datasets, comprising a total of 9,548 mammogram images. We performed extensive experiments showing that our method outperformed several state-of-the-art methods and that our method can enhance safer breast cancer diagnosis against adversarial attacks in clinical settings. 
\end{abstract}

\section{Introduction}
\begin{figure}[t!]
\centering
\includegraphics[width=0.8\linewidth, scale=0.5]{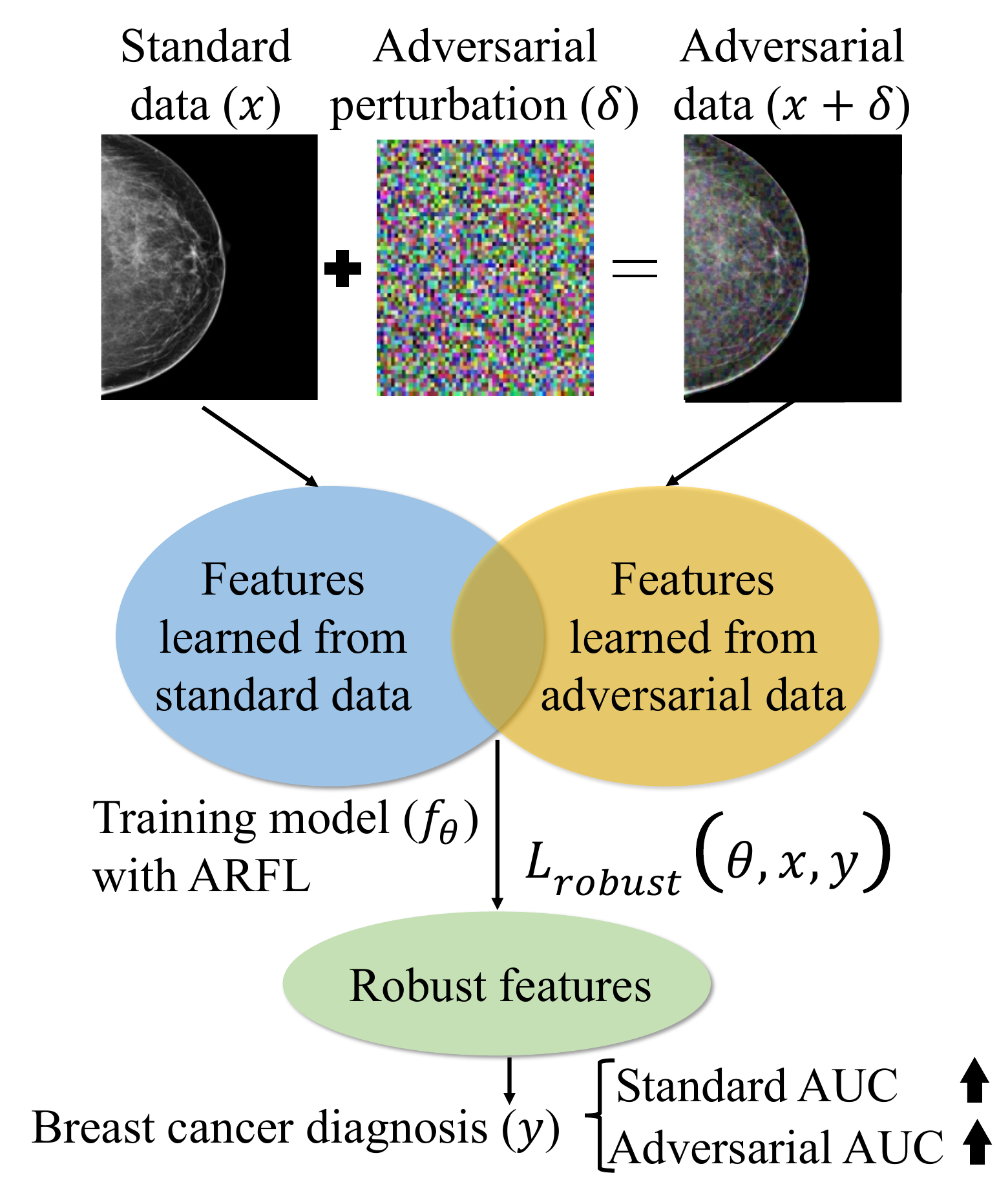}
\vskip -0.2in
\caption{We propose a novel method, adversarially robust feature learning (ARFL), which employs \( \mathbf{L}_{\mathbf{robust}}(\mathbf{\theta}, \mathbf{x}, \mathbf{y}) \) regularization to learn classification features that are robust on both standard and adversarial data to enhance breast cancer diagnosis.}
\label{fig:arfl-method}
\vskip -0.2in
\end{figure}
Adversarial samples can fool a deep learning classification model, where small and intentional perturbations may lead to unexpected results \cite{szegedy2013intriguing}. Adversarial attacking methods, such as projected gradient descent (PGD) \cite{madry2017towards} and fast gradient sign method (FGSM) \cite{goodfellow2014explaining}, have shown success on attacking classification of natural view images. Adversarial attacks also pose threats to deep learning-based medical applications, such as inducing unsafe diagnosis, fraudulent insurance claims, biased clinical trial outcomes, etc. \cite{finlayson2019adversarial}. In the medical imaging domain, previous studies showed adversarial samples may downgrade a model’s performance, as observed in image classification, detection, and segmentation  \cite{paschali2018generalizability, qi2021stabilized}. It is critical to develop deep learning models that are resistant to adversarial samples/attacks in order to deliver safe artificial intelligence (AI)-enabled medical applications.

Adversarial training, which trains a model by using a set of adversarially generated samples, is one of the few approaches to defend adversarial attacks \cite{shafahi2019adversarial}. Studies showed that by using the minimax optimization, adversarial training can improve a model’s adversarial robustness \cite{madry2017towards}. Adversarial samples may also serve as a special type of data augmentation to increase a model’s performance on the standard data (i.e., original clean data without adversarial perturbations) \cite{xie2020adversarial}. In the medical imaging domain, adversarial training-based methods have shown improved image diagnosis performance on either standard data \cite{han2021advancing} or adversarial data \cite{joel2022using}. However, it remains challenging for a model to maintain stable performance simultaneously on both the standard data and adversarial data \cite{picot2022adversarial, raghunathan2019adversarial, tsipras2018robustness, zhang2019theoretically, ilyas2019adversarial}. A previous study \cite{lin2020dual} indicated that the lack of exploiting the underlying manifold of data may be a key reason for this challenge.

While adversarial training has the benefits of resisting adversarial attacks, previous theoretical studies \cite{raghunathan2019adversarial, tsipras2018robustness} showed that adversarial training at the same time may make a model’s performance drop on standard data, which is undesirable, as it is equally important to maintain the model performance on both standard data and adversarial data \cite{picot2022adversarial, zhang2019theoretically}. A recent study showed that adversarial training could result in even worse results when training with limited data \cite{clarysse2022why}. In order to have a model perform stably on both standard data and adversarial data, one common approach is to directly merge the standard and adversarial datasets as a single set for training \cite{szegedy2013intriguing}. This, however, may not always work especially when the distributions/proportion of the two dataset exhibit considerable data shift. Researchers have considered standard data and adversarial data as two different domains to learn domain-invariant representations \cite{song2018improving}. Another approach, as proposed in a recent work \cite{chang2019domain}, is to perform training with separated batch normalization layers for standard data and adversarial data. Since the testing data’s distribution is usually unknown in priori, it is difficult for this approach to choose which batch normalization layer to use. Another method, TRADES \cite{zhang2019theoretically}, demonstrates there may be a theoretical trade-off of the performance between standard and adversarial data \cite{zhang2019theoretically}. With additional unlabeled data, robust self-training (RST) is shown capable of improving performance on adversarial data without sacrificing performance on standard data \cite{raghunathan2020understanding}. Overall, it remains an open research question in developing effective training strategies/methods to reconcile model performance on standard data and adversarial data.

In this study, we proposed a novel regularization method to regulate more effective learning of adversarially robust imaging features that are essential for both standard data and adversarial data (Figure 1). Specifically, our approach is to incorporate a feature correlation measure as an objective function to facilitate the training process, towards encouraging robust features and discouraging spurious features (those with low correlations with ground truth labels), when learning from a mix of standard and adversarial data. We name our method ARFL (Adversarially Robust Feature Learning). We implemented ARFL first on a synthetic two-moon benchmark dataset to prove the concept and then on two real-world and independent digital mammogram datasets for breast cancer diagnosis. We compared the effects of model training with and without the integration of ARFL. Also, we examined the learning effects of ARFL when mixing the standard data with adversarial data at a varying range of the ratio. In addition, we compared our method to several related methods, i.e., the batch normalization-based method \cite{chang2019domain}, TRADES \cite{zhang2019theoretically}, and multi-instance robust self-training (MIRST) \cite{sun2022mirst}. Extensive experiment results on the three datasets showed the clear benefits of ARFL in maintaining the model’s performance on both the standard data and adversarial data, and that our method outperformed the compared methods.

Our main contributions include the following aspects:
1). We proposed a novel regularization method, ARFL, to facilitate adversarial training to learn adversarially robust features from both standard data and adversarial data, and showed it outperformed the compared methods.
2). We showed our proposed method work promisingly on both the synthetic dataset and real-world imaging datasets (a total of 9,548 images) for breast cancer diagnosis, showing clear benefits of our method in medical applications.
3). We gained new insights about the optimal ratios of mixing standard data and adversarial data in impacting the effects of adversarial training.

\begin{figure*}[ht]
\centering
\includegraphics[width=\textwidth]{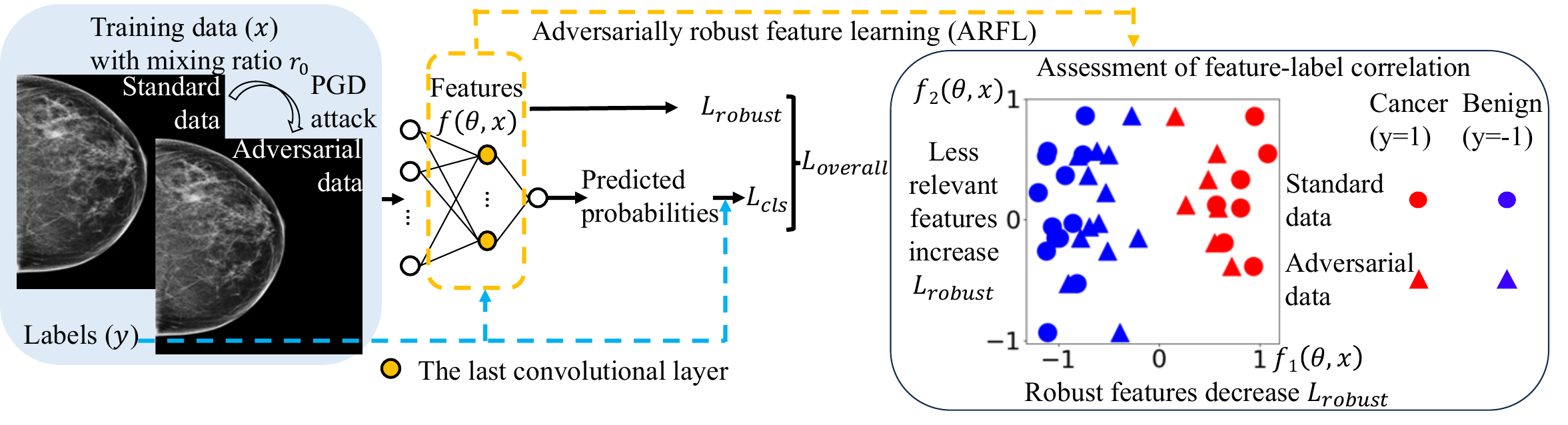}
\vskip -0.2in
\caption{Overview of the Adversarially Robust Feature Learning (ARFL) framework for breast cancer diagnosis. This figure illustrates the ARFL architecture utilizing both standard and adversarial mammographic data as inputs (\( \mathbf{x} \)). The adversarial training with ARFL focuses on extracting robust features \( \mathbf{f}(\mathbf{\theta}, \mathbf{x}) \) for computing the robust loss \( \mathbf{L}_{\mathbf{robust}} \). The ARFL approach is designed to enhance the identification of features that are robust on both standard and adversarial data while minimizing the influence of less relevant features.}
\label{fig:arfl-framework}
\vskip -0.2in
\end{figure*}
\section{Related Work}
While this paper is broadly related to AI model robustness and breast cancer, we focus on a brief review of the related work in two areas: AI safety studies for breast cancer diagnosis and adversarial defense methods.
\subsection{Safe AI for breast cancer diagnosis}
AI has shown promise and early success in enhancing various tasks for medical image analysis, including detection, classification, segmentation, reconstruction, registration, etc. \cite{chen2022recent}. AI-based breast cancer diagnosis models are under active development and clinical translation \cite{lotter2021robust}. It is imperative to ensure the deployment of such AI models are safe to patients, secure to clinical environments, and resilient to adversarial samples/attacks.

Adversarial security of AI models has attracted attention in the medical domain \cite{han2021advancing, mirsky2019ctgan, yao2020missthepoint}. Such studies on breast cancer/imaging is scarce, but more challenging, as malignancy information in breast imaging may be more subtle and heterogeneous \cite{kim2017breast}. Researchers showed that adversarial mammogram images produced by generative adversarial networks can fool both a breast cancer diagnosis model and experienced radiologists \cite{zhou2021machine}. A method called MIRST was introduced to defend untargeted adversarial attacks on breast ultrasound images \cite{sun2022mirst}.

\subsection{Adversarial defense}
In general, methods related to adversarial defense in medical image analysis can be divided into five categories: adversarial training, adversarial detection, image-level preprocessing, feature enhancement, and knowledge distillation \cite{dong2023adversarial}. A thorough review of the relevant literature can be found in a recent survey \cite{dong2023adversarial}. Here we focus on adversarial training and feature enhancement, as they are most closely related to our method.

Adversarial training improves adversarial robustness by incorporating adversarial data into the training process \cite{kurakin2016adversarial}. Typically, adversarial training is implemented by applying the minimax optimization to the conventional loss function in the standard training, i.e., searching for the worst-case perturbation in the inner maximization, while optimizing the model parameters in the outer minimization \cite{qian2022survey}. A previous study analyzed the minimax optimization and showed the effects of adversarial training as a powerful regularizer in improving adversarial robustness \cite{lyu2015unified}. Yet the minmax optimization commonly used in adversarial training does not consider the fitting to standard data, thus leading to a lower performance on standard data. Dual adversarial training \cite{joel2022using} remedies this limitation by training using both standard and adversarial data. Yet, it did not address the issue of how to assign the ratio of standard and adversarial data during training. The TRADES method \cite{zhang2019theoretically} introduces a regularization of model outputs and true labels as an additional objective alongside the minimax optimization, achieving a balanced and trade-off performance on both standard and adversarial data. Meanwhile, some studies suggest this trade-off effect may diminish when using adequate unlabeled data for self-training \cite{raghunathan2020understanding} or for virtual adversarial learning \cite{miyato2018virtual}. However, when the sample sizes of datasets are limited, it may be challenging for these methods \cite{raghunathan2020understanding, miyato2018virtual} to perform effectively. The strategy of our proposed method is different, as it does not rely on the use of additional data, but focuses on a novel mechanism to facilitate learning of robust features to improve model performance on both standard and adversarial data.

With regards to the feature enhancement approach, studies showed that adversarial training can enable the learning of features that are more consistent with human perception, i.e., features with strong correlation with the true labels \cite{tsipras2018robustness, ilyas2019adversarial}. For instance, one study \cite{allen2021feature} demonstrated that FGSM-based adversarial training can purify the features learned through standard training, like making some of the edge features more pure \cite{allen2021feature}. In another study, it was observed that PGD-based adversarial training was more effective at capturing long-range correlations, such as shapes and edges, and was less vulnerable to texture distortions, compared to the standard training methods \cite{zhang2019interpreting}. These findings indicate that the features derived from standard training often include redundant elements that are less likely or unlikely to contribute to the classification of adversarial data. In addition, previous studies \cite{raghunathan2019adversarial, hu2023understanding} also found that some features learned through adversarial training may result in a lower performance on standard data, and those features were termed as non-robust features. This finding about non-robust features indicate that features learned from adversarial training may also be redundant or conflictive. In view of the limitations of current adversarial training methods in learning features that are mutually robust on both standard and adversarial data, we propose to develop a novel feature-space regularization approach to learn adversarially robust features on medical images.
\section{Methods}
\subsection{Adversarially Robust Feature Learning (ARFL)}
When training a classification model with both standard and adversarial data, the model simultaneously fits two potentially different distributions. As shown in Figure 2, to encourage the learning of useful features from the mixed input and to reduce the chances the model learns from spurious correlations between the training data and truth labels, we introduced a regularization term, called adversarially robust feature learning (thus the name ARFL). As pointed by a previous work \cite{ilyas2019adversarial}, a feature’s usefulness can be measured by the expectation of feature-label multiplication, i.e., \(\mathbb{E}_{(x,y) \sim \mathcal{D}}(f_{i,j}(\theta, x) \cdot y))\), and the feature is called \(\rho\)-useful if the expectation is greater than \(\rho\). Inspired by such a correlation measurement, we designed a new loss function, named robust loss (denoted by \(L_{\text{robust}}\)), to characterize the feature-label correlation. \(L_{\text{robust}}\) is calculated by summing up the absolute values of the product of each feature and label over the feature map, as shown in Eq. 1.
\newcommand{\eq}[1]{\begin{equation} #1 \end{equation}}
\eq{
    L_{\text{robust}}(\theta, x, y)=-\frac{1}{HW} \sum_{i=1}^{H} \sum_{j=1}^{W}\sigma(\text{abs}(f_{i,j}(\theta, x)\cdot y))
}
where input \( x \) can be either a standard input with an underlying distribution of \( \mathbb{D} \); or an adversarial input from distribution \( \mathbb{D}' \); \( H \) and \( W \) respectively denote the width and height of a feature map of input \( x \); \(\text{abs}(\cdot)\) denotes the absolute value function; \( \sigma(\cdot) \) denotes the sigmoid function that scales the feature-label correlation; \( y \) denotes a positive or negative label \( \{\pm 1\} \); \( \theta \) denotes the model parameters; \( f_{i,j}(\theta, x) \) denotes the value of the feature map at position \((i,j)\). Considering that features near the output of a classification model contain more high-level information, we obtain the feature map from the last convolutional layer. \( L_{\text{robust}} \) encourages the model to learn features that are highly correlated with the labels. Different from the original method in \cite{ilyas2019adversarial}, we revised the method to measure useful features by adding an absolute-value operation to take into account both positive and negative correlations, and we also incorporated a sigmoid function to squash extreme loss values. Our method is appropriate as features showing either low positive correlations (yielding \(\rho>f_{i,j}(\theta, x)\cdot y>0 \)) or low negative correlations (yielding \(-\rho<f_{i,j}(\theta, x)\cdot y<0 \)) tend to be potentially less robust, leading to higher \( L_{\text{robust}} \) values. Then we integrate the adversarial loss and the robust loss as an overall loss for standard data as expressed in Eq. 2.
\eq{
    L_{\text{overall}}(\theta, x, y) = L_{\text{cls}}(\theta, x, y) + \lambda \cdot L_{\text{robust}}(\theta, x, y)
}
where \( L_{\text{cls}} \) denotes the binary cross entropy loss for binary classification tasks and \( \lambda \) is a weighting factor controlling the two objectives, i.e., the cross-entropy loss \( L_{\text{cls}} \) and the robust loss \( L_{\text{robust}} \).  
\subsection{Integrating ARFL into minimax optimization}
To construct adversarial data, we introduced some degree of adversarial perturbation generated by PGD \cite{madry2017towards} for the breast cancer diagnosis experiment and FGSM \cite{goodfellow2014explaining} for the synthetic experiment to standard data (\(x\)). PGD generates adversarial perturbations by iteratively maximizing the perturbation towards the direction of changing the predicted output whereas FGSM only optimizes via a single step. In constructing white-box defenses to the adversarial attacks, adversarial training minimizes the empirical loss of fitting the adversarial data while maximizing the same loss for the generated adversarial samples, as shown in Eq. 3.
\eq{
    \min_{\theta} \mathbb{E}_{(x,y) \sim \mathcal{D}} \left[ \max_{\delta \in \Delta (X)} L_{\text{cls}}(\theta, x + \delta, y) \right]
}
where \( \delta \) denotes the perturbation imposed to \( x \) within the specified set of valid perturbations \( \Delta \), and \( y \) denotes the truth label.

With both standard data and adversarial data in each training batch, we minimize the empirical loss by fitting both the standard data and adversarial data. We introduce Eq. 4 as the implementation of the minimax optimization process.
\eq{
    \begin{aligned}
        \min_{\theta} \mathbb{E}_{(x,y) \sim \mathcal{D}} \bigg[ &  (1-r) \cdot \max_{\delta \in \Delta (X)} L_{\text{cls}}(\theta, x + \delta, y) \\
        &+ r \cdot L_{\text{cls}}(\theta, x, y)  \bigg]
    \end{aligned}
}
where \( r \) denotes the ratio of the amount of standard data relative to the total amount of the data (standard data plus adversarial data) in each training batch. After integrating ARFL into adversarial training, we propose Eq. 5 for the minimax optimization on both standard and adversarial data.
\eq{
    \begin{aligned}
        \min_{\theta} \mathbb{E}_{(x,y) \sim \mathcal{D}} \bigg[ &(1-r)  \big(\max_{\delta \in \Delta (X)} L_{\text{cls}}(\theta, x + \delta, y) \\
        &- \lambda \cdot L_{\text{robust}}(\theta, x + \delta, y) \big) \\
        &+ r \cdot L_{\text{overall}}(\theta, x, y) \bigg]
    \end{aligned}
}
where \( r \) can take various values in the range \([0, 1]\) to define different training schemes. The term \( L_{\text{overall}} \) is defined in Eq. 2.
\section{Experiments}
\subsection{Datasets}
Our study was approved by the Institutional Review Board. We used three datasets for experiments. First, we used a synthetic set of the well-known two-moon dataset, which is simple and thus straightforward to demonstrate our proposed method. Using the scikit-learn software \cite{pedregosa2011scikit}, we created an upper moon and a lower moon for classification, including 10,000 samples for training and another 1,000 samples for testing. We set a noise ratio of 0.2.

Next, we examine the effects of our method on two real-world mammogram imaging datasets for breast cancer diagnosis. The first dataset is from Institution A and the second is the publicly available Chinese Mammography Database (CMMD) \cite{cui2021chinese}. The Institution A dataset was collected from a cohort of 1,284 women who underwent full field digital mammography screening. Each patient had one digital mammogram exam with up to four images of the two breasts (left craniocaudal [CC] view, left mediolateral oblique [MLO] view, right CC view, and right MLO view). Based on biopsy results, there are 366 patients diagnosed with breast cancer and 918 benign/negative cases. There are a total of 4,346 images. The images were acquired by a Hologic Lorad Selenia mammography system. The CMMD dataset was collected from a cohort of 1,775 patients who underwent mammography examination with both CC and MLO views. Based on biopsy, 1,310 patients are diagnosed with breast cancer and 465 patients are benign/negative, and there are a total of 5,202 mammogram images. The imaging data were acquired by a GE Senographe DS mammography system. Using the two independent datasets, our target classification task is to perform computer-aided diagnosis of classifying breast cancer (i.e., malignancy) vs. benign/negative findings.
\subsection{Experiment Settings}
\textbf{Model structure and training settings:} In our experiments for breast cancer diagnosis, we chose the widely used VGG16 \cite{simonyan2014very} model pre-trained on ImageNet \cite{deng2009imagenet} as the backbone of the classification model. We finetuned the fully connected layers and the last convolutional layer for binary classification. We assigned parameter \( r \) (Eq. 5) to 1, 0, and 0.5, respectively, to implement three training settings: 1) standard training (i.e., using only standard data), 2) adversarial training \cite{madry2017towards} (i.e., using only adversarial data), and 3) dual adversarial training \cite{joel2022using} (i.e., using 1:1 ratio of standard and adversarial data). We performed training with and without ARFL and compared the corresponding results under the three training settings. \( L_{\text{robust}} \)'s weight \( \lambda \) was set to 10 by experiments (See Appendix for the robustness analyses of parameter \( \lambda \) ).  On both mammographic datasets, we trained each model with 100 epochs.

In our experiments on the two-moon dataset, we employed a multi-layer perceptron architecture comprising layers with 2, 10, 10, and 1 node, respectively. The three training settings were configured in the same way as the experiments for breast cancer diagnosis. Features from the penultimate layer were used for calculation of \( L_{\text{robust}} \). We set \( \lambda \) to 0.5 by experiments and trained the model with 5000 epochs.

\textbf{Adversarial sample generation:} For the illustrative experiments on the synthetic two-moon dataset, we aim to demonstrate ARFL through against a stronger extent of the attacks. Therefore, we generated adversarial samples using the FGSM algorithm where we set the adversarial perturbation budget \( \varepsilon_1 \) to 0.05 and the attacking perturbation budget \( \varepsilon_2 \) to 0.2. For experiments on the breast cancer diagnosis, our goal is to mimic real-world defense and attacks. To generate adversarial samples by the PGD algorithm, we set the number of iterative steps to seven and the adversarial perturbation budget \( \varepsilon_1 \) to 0.01, which restricted the attacks within a \( \mathcal{L}_\infty \)-ball of size \( \varepsilon_1 \). This aligns with the defending perturbation budget used in previous works \cite{ma2021understanding, rice2020overfitting}. We provide an additional robustness evaluation of parameter \( \varepsilon_1 \) in the Appendix. The attacking perturbation budget \( \varepsilon_2 \) was set to 1e-4. We chose a smaller value of \( \varepsilon_2 \) with the purpose of generating visually imperceptible perturbations to simulate real-world attacking (because otherwise the obvious and large perturbations would be easily recognized by human visual observation).

\textbf{Comparison with related methods:} We compared our method to three related methods, including domain-specific batch normalization (DSBN) \cite{chang2019domain}, trading adversarial robustness off against accuracy (TRADES) \cite{zhang2019theoretically}, and multi-instance robust self-training (MIRST) \cite{sun2022mirst}. DSBN is a domain adaptation technique that allocates domain-specific affine parameters for data from different domains. DSBN was tested for adversarial training with standard data and adversarial data perturbed by the FGSM algorithm \cite{han2021advancing}. We replaced FGSM \cite{goodfellow2014explaining} with PGD \cite{madry2017towards}, aiming to measure our method's resilience against this more threatening challenge. TRADES is an adversarial defense method that balances model performance on adversarial data and standard data using KL-divergence for regularization. MIRST uses different levels of perturbations to generate adversarial examples as additional data for self-training. Note that DSBN was excluded from evaluation on the synthetic dataset, as it used a customized design of the convolutional networks thus it is incompatible to our simpler network architecture.

\textbf{Analysis of the mixing ratio parameter:} We performed a secondary experiment to examine the effects of mixing standard data with adversarial data at varying ratios (i.e., robustness analysis of parameter \( r \) in Eq. 5). While in dual adversarial training where \( r \) is set to 0.5, it is interesting to examine whether other values of this ratio may lead to different performance. In this experiment, we measured the diagnosis model’s performance additionally at \( r = 0.25 \) and \( r = 0.75 \) and compared to the effects when \( r = 0.5 \).

\textbf{Performance metric and statistical significance:} For the two-moon dataset, we replicated the experiment with five distinct random seeds and evaluated the model's performance in terms of accuracy. For the breast cancer diagnosis experiments, we measured the diagnosis model’s performance under five-fold cross validation, using the area under the receiver operating characteristics curve (AUC) of the binary classification as the performance metric. The AUCs of different methods were examined for statistical significance using p values measured by the Mann-Whitney U test.

\textbf{Visual assessment:} On the synthetic two-moon dataset, we visualized the decision boundaries of dual adversarial training and evaluated the impact of incorporating ARFL to distinguish the two classes. In order to visually assess the effects of feature learning using ARFL in the breast cancer diagnosis task, we plotted feature saliency maps on mammogram images. Feature saliency maps are calculated as gradients of loss with respect to the input.
\section{Results}
\subsection{Results on the Synthetic Dataset}
Table 1 presents the mean test set accuracy for both standard data (standard accuracy) and adversarial data (adversarial accuracy), along with standard deviations (std) and the average of standard and adversarial accuracies (mean accuracy), when using the two-moon dataset. As can be seen, while standard training achieves a high standard accuracy, its adversarial accuracy remains low (row A). Conversely, adversarial training boosts the adversarial accuracy, but at the cost of lowering the standard accuracy (row C). Dual adversarial training shows a modest improvement in both standard and adversarial accuracies compared to standard training (row E). The implementation of ARFL across the three training schemes resulted in slight increases in mean accuracies for both standard training (row B) and adversarial training (row D), along with a significant improvement in dual adversarial training (row F), surpassing the performance of both TRADES \cite{zhang2019theoretically} (row G) and MIRST \cite{sun2022mirst} (row H).
\begin{table*}[ht]
\centering
\caption{Model performance comparisons on the two-moon dataset. ($\uparrow$ - the higher the better)}
\label{table:two-moon-dataset}
\begin{tabular}{|l|c|c|c|}
\hline
\textbf{Training Method} & \textbf{Standard accuracy} & \textbf{Adversarial accuracy} & \textbf{Mean accuracy} \\ \hline
A. Standard training & 95.2 (0.4) & 71.4 (1.8) & 83.3 \\ 
B. Standard training + ARFL & 93.8 (2.5) & 73.0 (2.1) & 83.4 \\ 
C. Adversarial training \cite{madry2017towards} & 84.4 (3.4) & 81.8 (2.3) & 83.1 \\ 
D. Adversarial training \cite{madry2017towards} + ARFL & 84.8 (4.3) & 81.8 (0.8) & 83.3 \\ 
E. Dual adversarial training \cite{joel2022using} & 95.8 (1.2) & 72.6 (1.2) & 84.2 \\ 
F. Dual adversarial training \cite{joel2022using} + ARFL & 96.0 (0.9) & 79.6 (5.3) & 87.8 \\ 
G. TRADES \cite{zhang2019theoretically} & 95.6 (0.8) & 77.2 (0.7) & 86.4 \\ 
H. MIRST \cite{sun2022mirst} & 92.0 (0.6) & 76.4 (1.4) & 84.2 \\ \hline
\end{tabular}
\vskip -0.2in
\end{table*}
\begin{table*}[ht]
\centering
\caption{Model performance comparisons on the Institution A dataset. ($\uparrow$ - the higher the better)}
\label{table:institution-a-dataset}
\begin{tabular}{|l|c|c|c|}
\hline
\textbf{Training Method} & \textbf{Standard AUC} & \textbf{Adversarial AUC} & \textbf{Mean AUC} \\ \hline
A. Standard training & 69.2 (1.1) & 58.8 (1.4) & 64.0 \\ 
B. Standard training + ARFL & 70.0 (1.9) & 58.3 (3.5) & 64.2 \\ 
C. Adversarial training \cite{madry2017towards} & 61.7 (4.0) & 56.9 (5.3) & 59.3 \\ 
D. Adversarial training \cite{madry2017towards} + ARFL & 62.5 (4.3) & 59.2 (4.0) & 60.9 \\ 
E. Dual adversarial training \cite{joel2022using} & 65.7 (5.9) & 59.6 (9.4) & 62.7 \\ 
F. Dual adversarial training \cite{joel2022using} + ARFL & 69.3 (2.3) & 67.8 (2.4) & 68.6 \\ 
G. DSBN \cite{chang2019domain} & 54.1 (8.5) & 54.7 (9.0) & 54.4 \\ 
H. TRADES \cite{zhang2019theoretically} & 63.7 (3.5) & 63.2 (3.5) & 63.4 \\ 
I. MIRST \cite{sun2022mirst} & 63.0 (1.9) & 63.6 (1.7) & 63.3 \\ \hline
\end{tabular}
\vskip -0.2in
\end{table*}
\begin{table*}[ht]
\centering
\caption{Model performance comparisons on the CMMD dataset. ($\uparrow$ - the higher the better)}
\label{table:cmmd-dataset}
\begin{tabular}{|l|c|c|c|}
\hline
\textbf{Training Method} & \textbf{Standard AUC} & \textbf{Adversarial AUC} & \textbf{Mean AUC} \\ \hline
A. Standard training & 64.9 (4.2) & 41.5 (3.7) & 53.2 \\ 
B. Standard training + ARFL & 64.9 (4.4) & 41.5 (4.2) & 53.2 \\ 
C. Adversarial training \cite{madry2017towards} & 45.5 (4.6) & 43.7 (4.7) & 44.6 \\ 
D. Adversarial training \cite{madry2017towards} + ARFL & 48.6 (4.5) & 45.7 (4.7) & 47.15 \\ 
E. Dual adversarial training \cite{joel2022using} & 67.8 (3.3) & 66.3 (3.3) & 67.0 \\ 
F. Dual adversarial training \cite{joel2022using} + ARFL & 68.8 (3.3) & 67.3 (3.4) & 68.1 \\ 
G. DSBN \cite{chang2019domain} & 54.7 (6.9) & 55.5 (2.7) & 55.1 \\ 
H. TRADES \cite{zhang2019theoretically} & 64.8 (5.0) & 61.9 (5.1) & 63.4 \\ 
I. MIRST \cite{sun2022mirst} & 64.4 (2.6) & 64.8 (2.8) & 64.6 \\ \hline
\end{tabular}
\vskip -0.2in
\end{table*}

Figure 3 illustrates the comparisons of the decision boundaries of the dual adversarial training scheme and the effects after applying ARFL on the two-moon dataset. Comparing Figures 3A vs. 3C, as well as 3B vs. 3D, it clearly shows that the use of ARFL significantly reduces misclassifications, as marked by the black arrows in Figure 3. The implementation of ARFL adjusted the decision boundaries from an L-shape to an S-shape, as seen in Figure 3C and 3D. The S-shaped boundaries align more closely with the dual half-circle shape of the two-moon dataset. This observation verifies that ARFL can effectively capture the imaging patterns intrinsic to the dataset.
\begin{figure}[ht]
\centering
\includegraphics[width=\linewidth]{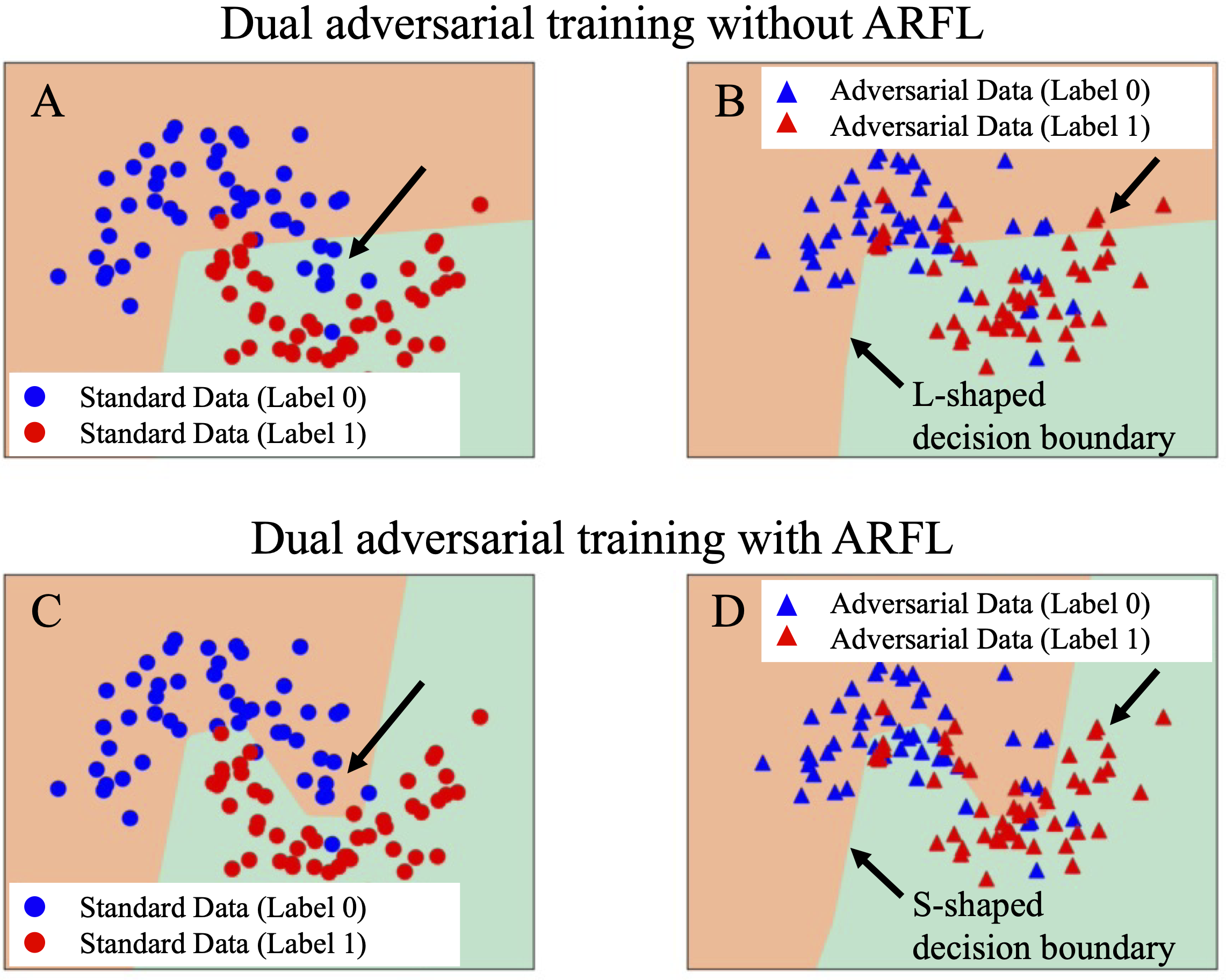}
\caption{Comparisons of adversarially robust feature learning (ARFL)'s effect on dual adversarial training \cite{joel2022using} on the two-moon dataset. The subfigures represent results from a multilayer perceptron subjected to the respective training settings, without ARFL (first row, i.e., Figure 3A and 3B) and with ARFL (second row, i.e., Figure 3C and 3D). The left column shows 50 samples randomly selected from standard test set, while the right column shows 50 samples randomly selected from the adversarial test set. ARFL is able to improve the shape learning of the latent data distribution, as seen from comparing the decision boundaries of Figures 3A vs. 3C or Figures 3B vs. 3D.}
\vskip -0.1in
\label{fig:arfl-comparison}
\end{figure}
\subsection{Results on Real-World Mammogram Datasets}
Table 2 and Table 3 show the mean AUC values and standard deviations on the test set of standard data and the test set of adversarial data, when using the Institution A dataset and CMMD dataset, respectively. As can be seen in Table 2, adversarial test had a substantially dropped performance under standard training (row A), which is the expected behavior for a standard model when facing adversarial attacks. When the model is trained by adversarial training (row C), adversarial test performance increased but at the same time the model downgraded in standard test - this sacrifice is undesirable for the slight benefit of adversarial robustness. When using dual adversarial training (row F), model performance largely increased in both standard test and adversarial test, showing the efficacy of this training method.

In terms of the benefits of ARFL, as shown in rows B, D, and F, while ARFL did not make a change in standard training (this is expected as ARFL is designed to mainly account for the mix of standard and adversarial data), it largely improved the performance for adversarial training (row D) and dual adversarial training (row F; here the benefits are the highest), showing the usefulness of our proposed method, in not only resisting adversarial attacks but also maintaining the performance in the original standard data. In the comparison, DSBN (row G), TRADES (row H), and MIRST (row I) exhibited lower performance compared to dual adversarial training with ARFL (row F). The underperformance of DSBN can be attributed to its limitation in selecting specific batch normalizations for test sets. Furthermore, this comparison highlights that ARFL's approach of regularizing through feature-label correlation is more robust than TRADES, which regularizes with prediction-label correlation. It also demonstrates ARFL can learn robust features without using multiple instances as MIRST does.

When examining the results of CMMD shown in Table 3, a very similar overall performance pattern is observed as seen in Table 2, which further verifies the efficacy and generalizability of our proposed method on an independent dataset. The dual adversarial training with ARFL also outperformed DSBN, TRADES, and MIRST. In addition, on both datasets, the AUCs of the dual adversarial training with ARFL are significantly higher (all \( p < 0.05 \)) than the AUCs of the adversarial training with ARFL.

It is worth mentioning that in Table 3 we noticed the adversarial training (row C) did not improve adversarial AUC compared to standard training (row A), though the standard deviation of the AUCs is also larger in row C compared to row A, showing the data heterogeneity may be higher in the CMMD dataset and that may lead to what we observed. Also note that the improvement resulted from adversarial training is also modest under adversarial test on the Institution A dataset (Table 2, row C vs. row A). Previous studies showed that adversarial training may only improve adversarial AUCs under the use of a very large dataset \cite{clarysse2022why}. This may partly explain the slight improvement observed in our study as our data scale is relatively small compared to large datasets.

When ranking all methods using the value of the last column in Tables 1-3 (i.e., the mean accuracy in Table 1 and the mean AUC in Tables 2 and 3), we observe that dual adversarial training with ARFL consistently stays on top of all methods. This demonstrates the benefit of ARFL on reconciling the performance on standard data and the performance on adversarial data when training a model with both data. In comparison, ARFL provides modest improvement of mean accuracy/AUC to models with standard training or adversarial training. This comparison underscores ARFL’s effectiveness in adapting to unknown data distributions. In other words, ARFL enhances the model fitting to mixed data from two different distributions and without compromising the effects when fitting to data of a single distribution.

\subsection{Analysis of the Mixing Ratio Parameter}
Figure 4 shows the results of model performance when varying the mixing ratio (\( r \)) of standard data relative to the total data. In the Institution A dataset, it confirms that \( r = 0.5 \) is an optimal parameter value (i.e., dual adversarial training), as it achieved the highest model performance. In the CMMD dataset, while \( r = 0.5 \) is a good parameter value with high AUCs, the model achieved the highest performance at \( r = 0.75 \). This means that the reported model AUC values in Table 3 could go higher if \( r \) is set to be 0.75. While that, we reported results on both datasets with \( r = 0.5 \) for the consistency reason and also to make a fairer comparison with previous studies \cite{joel2022using, zhang2019theoretically, chang2019domain, sun2022mirst}. While dual adversarial training (\( r = 0.5 \)) has been the common practice in the related work, this experiment indicates that for specific datasets and classification tasks, a different mixing ratio of the standard data and adversarial data may lead to a potentially higher model performance, which merits further investigation in future work.
\begin{figure}[h]
\centering
\includegraphics[width=0.8\linewidth]{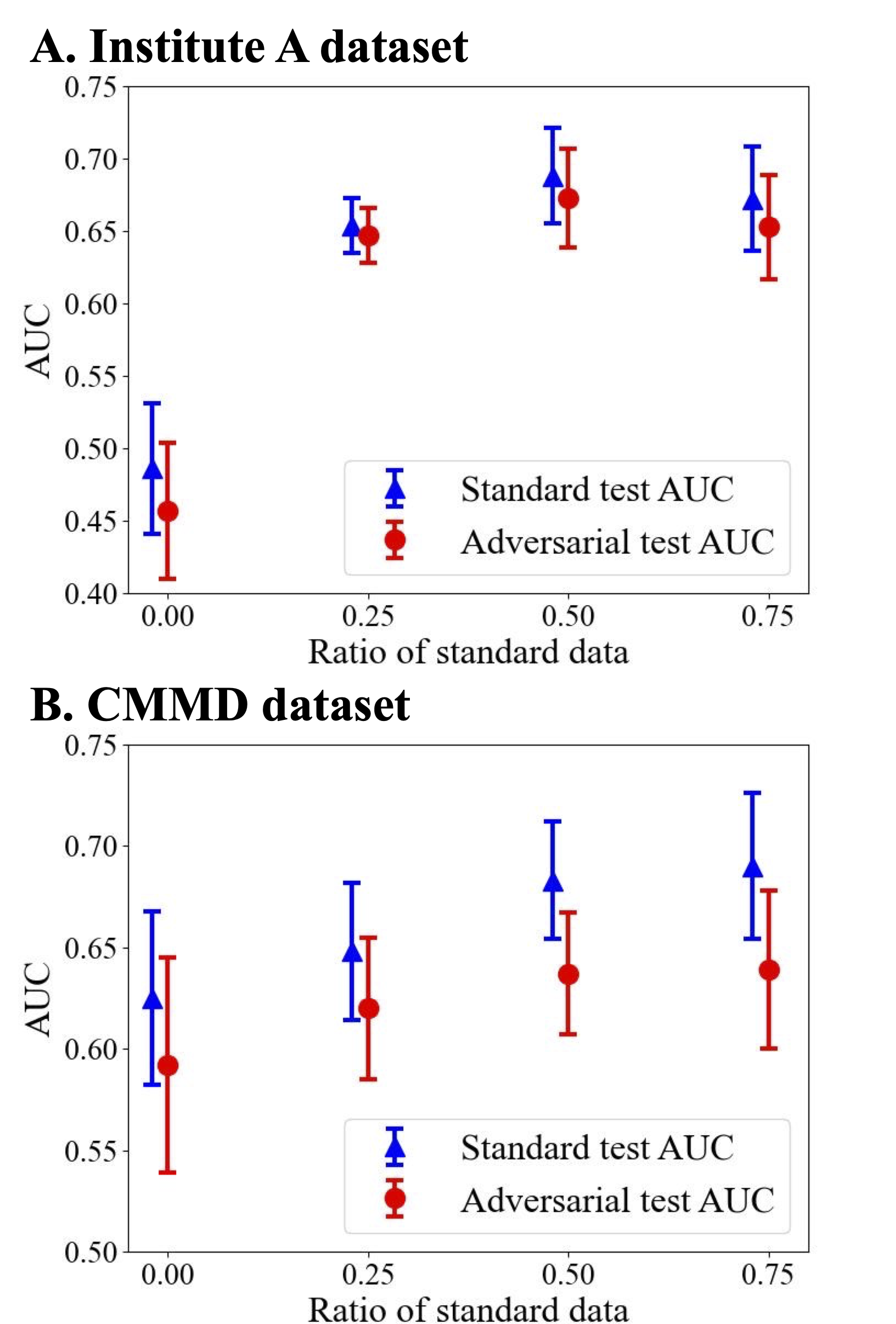}
\caption{Robustness analysis of parameter (the mixing ratio of standard data relative to total data). Shown are the AUC values with respect to varying values. Error bars reflect standard deviations. Note that at the same values, the display of the red markers and error bars are purposely shifted a little bit to the right, for better visualization (avoiding overlap).}
\label{fig:robustness-analysis}
\vskip -0.1in
\end{figure}
\subsection{Visual Assessment on Mammogram Images}
Figure 5 illustrates on example mammogram images the feature saliency maps for models trained with dual adversarial training with and without ARFL. In these maps, regions with sharp intensity contrast indicate important features, where higher gradients suggest stronger influence on the classification performance \cite{tsipras2018robustness}. The comparison shows that incorporating ARFL results in a greater number of sharply contrasted regions, suggesting that ARFL enhances the learning of discriminative imaging features for the diagnosis purposes. Note that we demonstrate the saliency maps mainly on standard data as these clean data are better cases to illustrate and perceive the effects.
\begin{figure}[h]
\centering
\includegraphics[width=\linewidth]{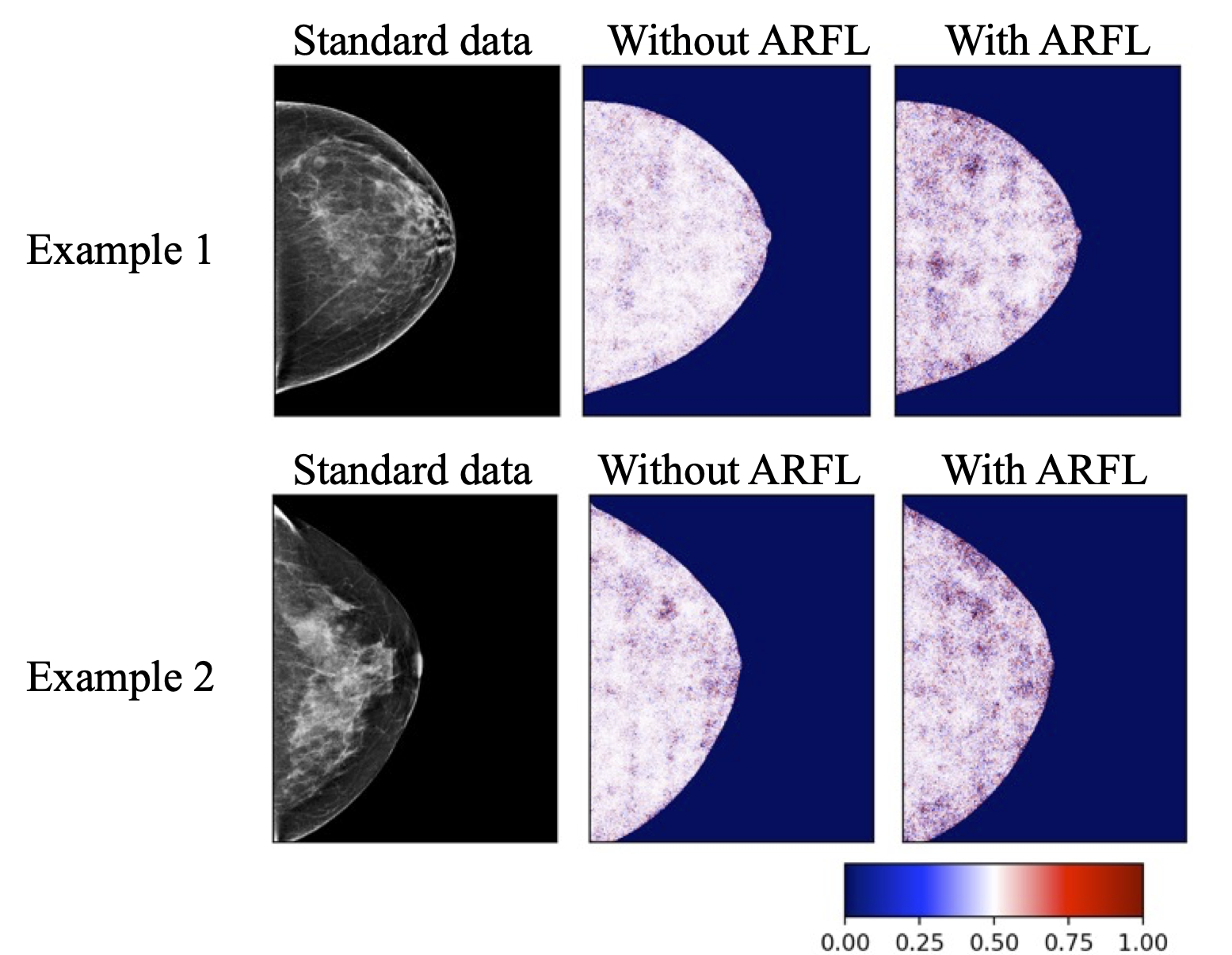}
\vskip -0.1in
\caption{Feature saliency maps of mammogram images from models trained using dual adversarial training with and without the integration of ARFL. The color bar represents the scaled gradients between zero and one. More regions with sharp contrast indicate more important features.}
\label{fig:feature-saliency}
\vskip -0.2in
\end{figure}
\section{Conclusion}
In this work, we designed a novel method, ARFL, to facilitate adversarially robust feature learning in adversarial training. ARFL facilitates the learning process towards identifying features that are strongly correlated with true labels. On both the synthetic data and breast imaging datasets, ARFL showed clear benefits in resisting adversarial attacks and maintaining stable model performance on standard data. Our extensive experiments on the three datasets showed similar effects and the generalizability of our method on the two clinical datasets from different sources. ARFL also outperformed the several compared methods. For future work, we will extend the evaluation of our method against other types of adversarial attacks.

\clearpage
\section*{Broader Impact}
AI-based medical diagnosis systems are being quickly translated to clinical settings. The safety and security of medical AI systems are at the highest concern to ensure patient safety. This study investigates a novel deep learning method and real-world medical application to enhance safety of breast cancer diagnosis models against adversarial attacks. This work contributes to advancing accuracy, reliability, and trustworthiness of AI-empowered medical diagnostic technologies to transform patient care.

\section*{Acknowledgment}
This work was supported by a National Science Foundation (NSF) grant (CICI: SIVD: \#2115082), a National Institutes of Health (NIH)/National Cancer Institute (NCI) grant (1R01CA218405), the grant 1R01EB032896 (and a Supplement grant 3R01EB032896-03S1) as part of the NSF/NIH Smart Health and Biomedical Research in the Era of Artificial Intelligence and Advanced Data Science Program, an Amazon Machine Learning Research Award, and the University of Pittsburgh Momentum Funds (a scaling grant) for the Pittsburgh Center for AI Innovation in Medical Imaging. This work used Bridges-2 at Pittsburgh Supercomputing Center through allocation [MED200006] from the Advanced Cyberinfrastructure Coordination Ecosystem: Services \& Support (ACCESS) program, which is supported by NSF grants \#2138259, \#2138286, \#2138307, \#2137603, and \#2138296.

\bibliography{example_paper}
\bibliographystyle{icml2024}

\clearpage
\newpage
\appendix
\section*{Appendix}
\section{Parameter Robustness Analyses}
In this section, we present two supplementary experiments conducted to evaluate the effects of the proposed Adversarially Robust Feature Learning (ARFL) method on two hyperparameters: the weighting factor (\( \lambda \)) and the adversarial perturbation budget (\( \varepsilon_1 \)). These experiments provide a more comprehensive assessment regarding the two hyperparameters' impact on the model's performance on standard data and adversarial data.
\subsection{Effects of Weighting Factor (\( \lambda \))}
The weighting factor \( \lambda \), which controls the influence of \( L_{\text{cls}} \) and \( L_{\text{robust}} \) in the model, was varied from 0.1 to 100.0. We applied ARFL in the context of dual adversarial training to determine the optimal balance point, where the model efficiently learns robust features without compromising classification performance.
\subsection{Effects of Adversarial Perturbation Budget (\( \varepsilon_1 \))}
We investigated the impact of varying the adversarial perturbation budget \( \varepsilon_1 \) within the range of 0.005 to 0.1. We used 0.1 as the upper bound considering literatures \cite{joel2022using,han2021advancing} and characteristics of mammogram images. Using the PGD method, we generated adversarial data constrained by this budget and incorporated the data into the adversarial training process. The aim was to observe how different levels of adversarial perturbation during adversarial training influence the model's defense against adversarial attacks.
\subsection{Experimental Setup}
For both experiments, we employed five-fold cross-validation on the Institution A dataset. The model's performance, in terms of the mean Area Under the Curve (AUC) and the standard deviation, was used to measure the model's performance at different settings. 
\subsection{Results}

Figure 6 shows the effects of adjusting the weighting factor, \( \lambda \), which regulates the balance between cross-entropy loss (\( L_{\text{cls}} \)) and robustness loss (\( L_{\text{robust}} \)). As the value of \( \lambda \) increases from 0.1 to 100.0, we observe an initial increase in the test AUC, followed by a decrease. The highest test AUC is achieved when \( \lambda \) equals to 10.0.

Figure 7 presents the model's test AUCs in response to varying the adversarial perturbation budget, \( \varepsilon_1 \). It is observed that as \( \varepsilon_1 \) increases, both the standard test AUC and the adversarial test AUC initially increase and then overall appear stabilized when \( \varepsilon_1 \) reaches 0.01 and further. This trend suggests an optimal range for \( \varepsilon_1 \) in producing adversarial data for adversarial training in our study/data.

The two experiments of parameter robustness supported the use of optimal parameter values in our main experiments. Note that for different datasets and/or medical tasks, the optimal parameter values may be different.
\begin{figure}[h]
\centering
\includegraphics[width=\linewidth]{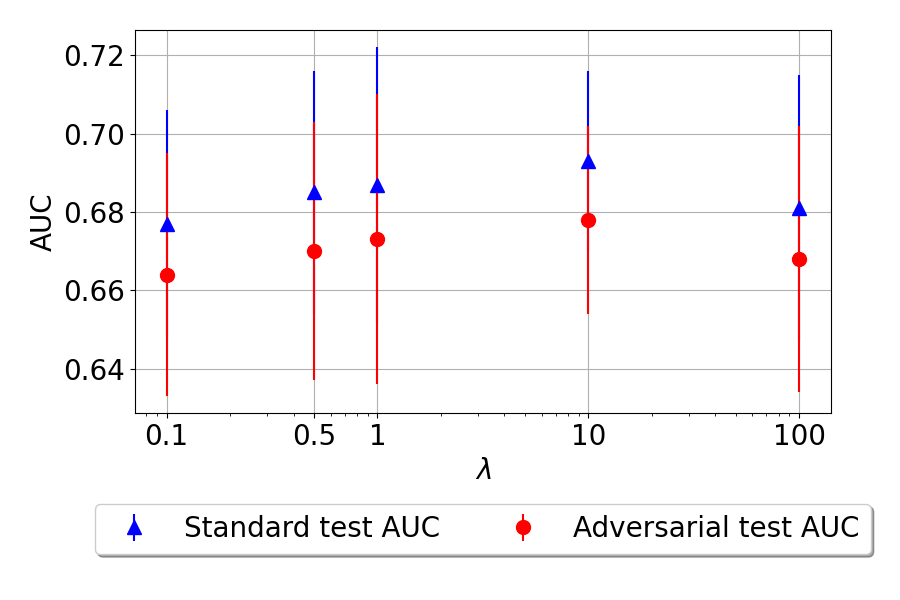}
\vskip -0.2in
\caption{
Evaluation of model performance (i.e., standard test AUC and adversarial test AUC) with respect to the weighting factor \( \lambda \), which balances the cross-entropy loss (\( L_{\text{cls}} \)) and the robustness loss (\( L_{\text{robust}} \)), on the test AUCs during dual adversarial training.
}
\label{fig:robustness-analysis}

\end{figure}

\begin{figure}[h]
\centering
\includegraphics[width=\linewidth]{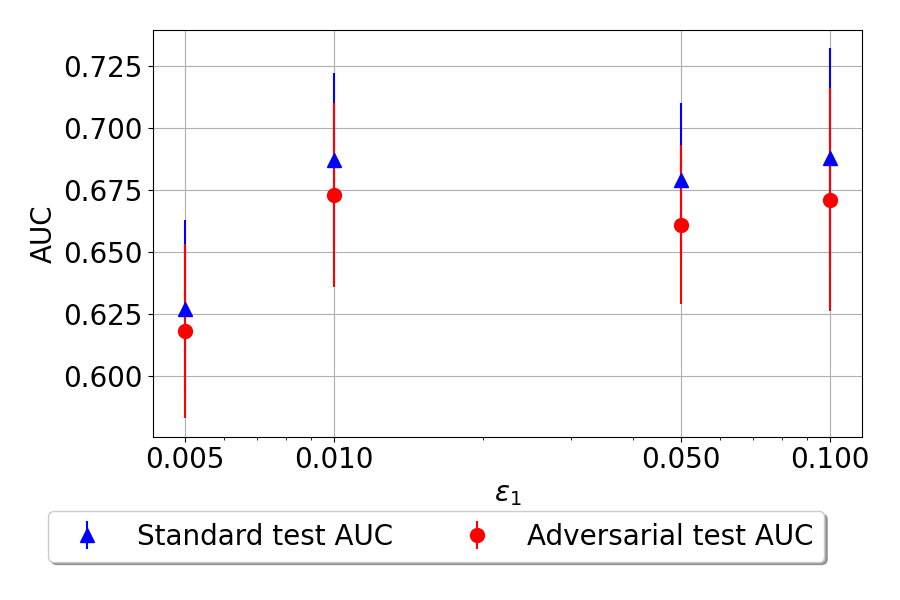}
\vskip -0.2in
\caption{
Variations of model performance (i.e., standard test AUC and adversarial test AUC) with respect to a range of the adversarial perturbation budget \( \varepsilon_1 \), which was used to generate adversarial data for dual adversarial training.}
\label{fig:epsilon}
\end{figure}

\section{Code}
The code repository of this study to reproduce the experiments will be shared upon publication of this work.
\section{Data}
The two imaging datasets are available for research, where the CMMD dataset is publicly available and can be downloaded from https://bit.ly/cmmd-dataset. The Institution A dataset is an internal private dataset and may be available to interested users upon request, after an approval from the institution along with a signed data use agreement and/or a material transfer agreement. The synthetic two-moon dataset can be generated using the code stored in the GitHub code repository. 


\end{document}